\documentclass[prl,showpacs,reprint,longbibliography]{revtex4-1}
\usepackage{amsmath}
\usepackage{epsfig}
\usepackage[outdir=./]{epstopdf}

\def\eq#1{Eq.~(\ref{#1})}

\def\fig#1{Fig.~\ref{#1}}

\begin{document}
	
	\title[Beats and Expansion of Two-Component Bose-Einstein Condensates]{Beats and Expansion of Two-Component Bose-Einstein Condensates\\ in the Thomas-Fermi Limit}
	
	\author{James Q. Quach} 

	\address{Institute for Solid State Physics, University of Tokyo, Kashiwa, Chiba 277-8581, Japan}
	\address{School of Physics, The University of Melbourne, Victoria 3010, Australia}
	
	\email{quach.james@gmail.com}
	
	\begin{abstract}
		A unique feature of multi-component BECs is the possibility of beating frequencies in collective oscillations. We analytically determine this beating frequency for the two-component BEC in one-dimension. We also show that the Thomas-Fermi approximation, where the quantum pressure is neglected, describes well the  expansion of the two-component condensate released from an harmonic trap.
	\end{abstract}
	
	
	\maketitle
	
	\section{Introduction}
	
	Two-component Bose-Einstein condensates (BECs) now form a rich area of theoretical and experimental  investigation. Two-component BECs have been realised with two different hyperfine spin states of $^{87}$Rb \cite{myatt97, hall98, maddaloni00, delannoy01, mertes07, anderson09, tojo10}, different atomic species  \cite{ferrari02, modugno02, thalhammer08, mccarron11}, and different isotopes of the same atomic species  \cite{papp08}. They have been shown to exhibit miscible and immiscible behaviour  \cite{papp08, hall98, tojo10}, modulation instabilities  \cite{robins01, ronen08}, dark-bright solitons \cite{busch01,hamner11}, and vortices  \cite{matthews99, feder99}. A unique feature of multi-component BECs, is the possibility of a beating frequencies; this has been seen in numerical investigations \cite{kasamatsu04} and for the special case of dark-dark solitons, even experimentally observed \cite{hoefer11}.
	
	This exhibition of a large range of interesting behaviour is the result of the numerous tunable parameters: atom number, mass, interaction strength, trapping frequency, and trap ellipticity for each species or component. In fact the vast parameter space means that it is difficult to fully investigate with computer simulations alone, and analytical solutions are needed. To make the equations tractable, the Thomas-Fermi (TF) approximation is used: in this approximation the kinetic term is either completely or partially ignored. Equations of motion (EOM) for two-components BECs have been developed using the TF approximation \cite{graham98,kasamatsu04}. In this work, we will show that such analytical equations can capture the beating frequencies of two-component BECs, which have previously been seen in numerical simulations but only qualitatively described. We will also confirm that these EOM are valid for the case of an expanding two-component BEC released from a trap.
	
	In Sec. 2 we present the coupled Gross-Pitaevskii equation (GPE) and its hydrodynamical form, and discuss the validity of the TF approximation. In Sec. 3 we consider the small oscillation regime to analytically calculate the beating frequencies of two-component BECs. In Sec. 4 we consider the large amplitude regime to analytically determine the dynamics of the expansion of the two-component BEC released from a trap. In both Sec 3. and 4. we will confirm the correctness of our analytical predictions with numerical results.

	\section{The hydrodynamical equations in the Thomas-Fermi limit}
	\label{sec:Model}
	
	In the limit of near-zero temperatures, the mean-field of the two-component BEC is well-described by the coupled Gross-Pitaevskii equation \cite{ho96}. We consider the extreme asymmetrical potential $V = m ({\omega^2_{x}x^2 + \omega_{y}^2y^2 + \omega_{z}^2z^2})/2$, where $\omega_y,\omega_z \gg \omega_x$. Numerous experiments with BECs in such extremely asymmetrical traps have been performed \cite{gorlitz01,moritz03,haller09}.  For convenience we will assume that the mass, $m$, number of atoms, $N$, and trapping frequencies, $\omega$, of the two components are identical, so that the variable parameter is the interaction matrix, $g=\begin{pmatrix}
	g_{11}& g_{12}\\	
	g_{21}& g_{22}
	\end{pmatrix}$. An effective one-dimensional (1D) GPE describes this system: 
	\begin{gather}
	i\frac{\partial\tilde{\psi}_{l,x}}{\partial \tilde{t}} = \bigl(-\frac{\partial^2}{2\partial \tilde{x}^2}+\tilde{\omega}(t)^2\frac{\tilde{x}^2}{2}+\tilde{g}_{ll}|\tilde{\psi}_{l,x}|^2+\tilde{g}_{ll'}|\tilde{\psi}_{l',x}|^2\bigr)\tilde{\psi}_{l,x}~,
	\label{eq:gpe1D}
	\end{gather}
	where $l\neq l'$ subscripts ($l=1,2$) mark the components. The tilde indicates we have used the typical rescaling of variables to get a dimensionless GPE \cite{cerimele00,bao03}: $\tilde{t} = \omega_x t$, $\tilde{x} = x/L$, $\tilde{\omega} = \frac{\omega_x(t)}{\omega_x(0)}$, $\tilde{\psi} = \sqrt{L} \psi$, $\int|\tilde{\psi_l}|^2 d\tilde{x} = 1$, where $L=\sqrt{\frac{\hbar}{m\omega_x}}$ is the harmonic oscillator length scale and $\psi_l$ the condensate wave function or order parameter of component $l$. As energy along the $y,z$-axis is much larger than along the $x$-axis, we make the approximation that there is no excitations in the $y,z$-direction. This leads to $\tilde{g}=\frac{\sqrt{\omega_y(0)\omega_z(0)}}{2\pi\omega_x(0)}g$~\cite{jackson98,dunjko01,bao03,ge05}. With this understanding, we will drop the use of the tilde and co-ordinate subscript.
	
\begin{figure}[tb]%
		\centering
		\includegraphics[width=\columnwidth]{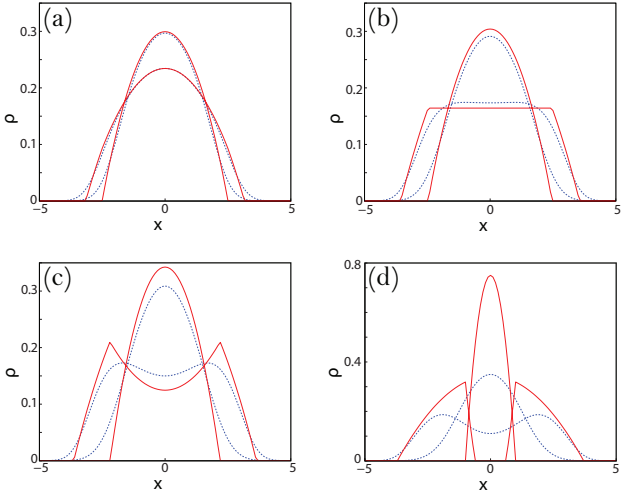}%
		\caption{(Color online) Some examples of 1D density profiles of two-component BECs in an harmonic trap ($\omega=1$) for $g_{12}^2/|g_{11} g_{22}|=\text{(a) }0.005,\text{(b) }0.5,\text{(c) }0.72,\text{(d) }0.98$. The solid (red) lines represent the TF approximation and the dotted (blue) lines were numerically calculated from the full coupled GPE. As $g_{12}^2/|g_{11} g_2| \rightarrow 1$, the TF approximation quickly worsens.}%
		\label{fig:TF}%
	\end{figure}

	When there are a lot of particles and the mean-field energy is large compared to the kinetic energy, the TF approximation, where the kinetic term of the GPE is ignored, has been shown to provide a good estimate of the ground state of the time-independent single-component GPE \cite{baym96}. The validity of the TF approximation in the two-component case, is further restricted to the regime $g_{12}<|g_{11}|,|g_{22}|$ \cite{corro09}. The groundstate in the TF limit is given by ($\rho \equiv |\psi|^2$)
	\begin{equation}
	\rho_l^{\text{g.s.}} = \rho_l^o H(\rho_l^o) H(\rho_{l'}^o) + \rho_l^s H(-\rho_{l'}^o) H(\rho_{l}^s)~,
	\label{eq:rho}
	\end{equation}
	where
	\begin{gather}
	\rho_l^o = \frac{g_{l'l'}(\mu_l - V_l) - g_{ll'} (\mu_{l'} - V_{l'})}{|g|}~,\\
	\rho_l^s = \frac{\mu_l - V_l}{_{ll}}~.
	\label{eq:rho_comp}
	\end{gather}
	$H$ is the Heaviside step function, and $\mu_l$ is the chemical potential of component $l$. The superscripts $o$ and $s$, indicate regions where the components overlap and are singular, respectively. \fig{fig:TF} compares groundstates under the TF approximation with groundstates numerically calculated from the full two-component GPE in a 1D harmonic trap: the TF approximation of the groundstate of the time-independent coupled GPE worsens as $g_{12}^2/|g_{11} g_{22}| \rightarrow 1$. The TF approximation can also yield asymmetrical groundstate solutions \cite{trippenbach00}, which we will not consider in this work.
	
	By writing the complex order parameter in terms of a density and a phase, $\psi_l (x,t) = \sqrt{\rho_l(x,t)} e^{i \phi_l (x,t)/\hbar}$, the GPE can be reformulated into a set of coupled hydrodynamical equations. In this formulation the kinetic component is split into a phase gradient dependent component ($m v_l^2/2$) and a density gradient dependent component, known as the \textit{quantum pressure} ($-\frac{\hbar^2}{2 m \rho_l}\frac{\partial^2\sqrt{\rho_l}}{\partial x^2}$); the velocity is defined as $v_l=\frac{1}{m}\frac{\partial\phi_l}{\partial x}$. As $v_l$ gives the velocity of the condensate flow, we will refer to $m v_l^2/2$ as the \textit{current energy}. When the number of particles is large the density profile becomes smooth, and the quantum pressure term can be neglected \cite{stringari96}. In this TF limit, the hydrodynamical equations for the two-component BEC are
	\begin{gather}
	-\frac{\partial\rho_l}{\partial t} = \frac{\partial ({v}_l \rho_l)}{\partial x}~,\label{eq:hydrodynamic1}\\
	-m_l \frac{\partial v_l}{\partial t} = \partial_x(V_l + g_{ll}\rho_l + g_{ll'} \rho_{l'} + \frac12 m_l v_l^2)~.\label{eq:hydrodynamic2}
	\end{gather}
	This hydrodynamical formulation of the GPE will prove useful in the analyses of the following sections.

	\section{Beat frequency}
	
	Investigations into collective oscillations were among the first experiments conducted following the realization of BECs. These excitations can be introduced with modulations in the natural frequency of the trapping potential \cite{jin96, mewes96, castin96, dalfovo97, garcia99a, garcia99b} or the $s$-wave scattering length \cite{pollack09, vidanovic11}. Investigations into two-component oscillations have been studied both analytically \cite{busch97,graham98,kasamatsu04} and numerically \cite{esry98,pu98,gordon98,kasamatsu04}. In particular, beating frequencies were seen in the numerical simulations of Ref. \cite{kasamatsu04}. The presence of these beating frequencies were only \emph{qualitatively} described in this work. In this section we will analytically quantify this beating frequency for the 1D case. 
	
	Small amplitude oscillations can be analysed by linearising the hydrodynamical equations [\eq{eq:hydrodynamic1} and (\ref{eq:hydrodynamic2})] around their equilibrium values: $\rho_l = \rho_l^{\text{g.s.}}+ \delta\rho_l$ and $v_l = \delta v_l$. We further make the  restriction that the two components of the BEC completely overlap (we will also show how the results degrade as we move away from this regime). This assumption means that this analysis is most valid for miscible systems~\cite{kasamatsu04}. The linearised hydrodynamical equations are,
	\begin{equation}
	\frac{\partial^2\delta\rho_l}{\partial t^2} = \frac{1}{m} \frac{d}{dx}[\rho_l^{\text{g.s.}}\frac{d}{dx}(g_{ll} \delta \rho_{l} + g_{ll'} \delta \rho_{l'})]~.
	\label{eq:linearised_eom}
	\end{equation}
	
	An analytical solution to \eq{eq:linearised_eom} is given by the ansatz $\delta\rho_l(x,t) = \alpha_{l,0}(t) - \alpha_l(t)x^2$, where the time dependences of $\alpha_{l,0}(t)$ are determined by the conditions $\int dx \delta\rho_{l}=0$ \cite{kasamatsu04}. One then immediately recognises \eq{eq:linearised_eom} as an equation for coupled oscillators,
	\begin{equation}
	\frac{d^2}{dt^2}
	\begin{pmatrix}
	\alpha_1\\
	\alpha_2
	\end{pmatrix}=
	\begin{pmatrix}
	g_{11}\kappa_1 & 	g_{12}\kappa_1\\
	g_{12}\kappa_2 & 	g_{22}\kappa_2
	\end{pmatrix}
	\begin{pmatrix}
	\alpha_1\\
	\alpha_2
	\end{pmatrix}
	\label{eq:linear_mtx}
	\end{equation}
	where $\kappa_l \equiv 3\omega^2(g_{l'l'} - g_{ll'} ^2)/|g|$. The solution to \eq{eq:linear_mtx} takes the form $\alpha_l = A_l e^{i\Omega t}$. Substitution of this solution into \eq{eq:linear_mtx} and solving the resulting eigenvalue equations give the normal mode frequencies $\Omega^- = \sqrt{3} \gamma \omega$ and $\Omega^+ = \sqrt{3}\omega$, where $\gamma \equiv \sqrt{(g_{ll}-g_{ll'})(g_{l'l'}-g_{ll'})/|g|}$. Note that as we are assuming positive interaction strengths, in the TF limit where $g_{12} \gg g_{11}, g_{22}$, $\gamma$ is upper bounded by 1. Analogous to the classical problem of two coupled oscillators, the difference in the normal mode frequencies give the beating frequency,
	\begin{align}
	f_b &= \frac{\Omega^+ - \Omega^-}{2}\\
	&= \frac{\sqrt{3}}{2}(1-\gamma)\omega~.
	\label{eq:f_b}
	\end{align}
	
\begin{figure}[tb]%
		\centering
		\includegraphics[width=\columnwidth]{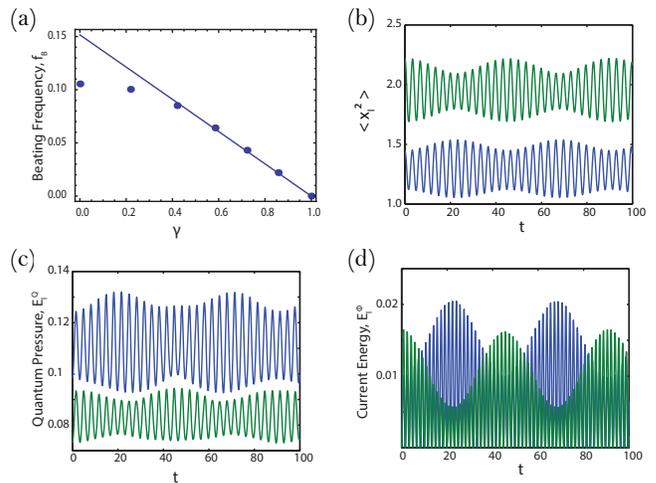}%
		\caption{(Color online) Beating frequencies of a two-component BEC in a 1D harmonic trap ($\omega = 1.1$). (a) The analytically predicted beating frequency [\eq{eq:f_b}] is given by the solid line. The results of computers simulations are plotted with dots. In its regime of applicability, i.e. when the two components overlap and their density distribution are approximately quadratic, \eq{eq:f_b} matches the numerical results well. These assumptions become increasing violated as $\gamma$ approaches zero, and \eq{eq:f_b} is no longer applicable. (b) plots the mean-squared displacement for $\gamma = 6/7$. It shows the typical beating feature of the two-component BEC. The beating arises from the transfer of kinetic energy between the two BEC components.  (c) Quantum pressure ($E_l^Q$) and (d) current energy ($E_l^\phi$) of the two BEC components given by the numerical simulation: the quantum pressure dominates the kinetic energy. In (b)-(d) blue represents component 1, and green component 2.}%
		\label{fig:pert}%
	\end{figure}

	In \fig{fig:pert} we compare \eq{eq:f_b} with the results of numerical simulations. In our simulation, the BEC is initially in the groundstate of a trapping potential with frequency $\omega(0)=1$. We excite the collective oscillation by perturbing the trapping potential, $\omega(t>0)=1.1$. \fig{fig:pert}(a) compares the analytical and numerically beating frequencies as a function of $\gamma$. It shows that \eq{eq:f_b} corresponds well with the numerical results when it is in its region of applicability i.e. when the two components overlap and their density distribution are approximately quadratic [e.g. \fig{fig:TF}(a)] . However as $\gamma$ approaches zero,
	these assumptions are violated [e.g. \fig{fig:TF}(b)], and \eq{eq:f_b} is no longer applicable. \fig{fig:pert}(b) plots the mean-squared displacement  [$<x_l^2> \equiv \int \rho_l(x,t) x^2 dx$] for $\gamma = 6/7$, showing the typical beating characteristic of the two-component BEC.
	
	The energy of the system is composed of an interaction and kinetic component. The interaction energy is given by $E_l^\text{int}=\int{(\frac{g_{ll}}{2} \rho_l^2 + \frac{g_{12}}{2} \rho_l \rho_{l'})}dx$. The kinetic energy, $E_l^\text{kin} = E_l^\phi + E_l^Q$, is composed of the current energy $E_l^\phi = \frac12 m_l <v_l^2>$ and the quantum pressure $E_l^Q=\int{\frac{1}{2m} |\partial_x\sqrt{\rho_l}|^2  dx}$. The TF approximation assumes that $E_l^Q=0$. \fig{fig:pert}(c) and (d) plots $E_l^Q$ and $E_l^\phi$ respectively. It is interesting to point out that the quantum pressure here plays the dominate role in the kinetic energy, drastically violating the TF assumption that $E_l^Q=0$. In spite of this crude approximation, the fact that \eq{eq:f_b} can accurately predict the beat frequency, speaks of the usefulness of the TF approximation.

	\section{Expansion of released BECs}
	
	The method used in the previous section is only valid for small amplitude oscillations. Following the method of Ref.~\cite{dalfovo97}, Ref.~\cite{kasamatsu04} developed EOM for two-component BECs which do not linearise the hydrodynamical equations. In this section we will confirm that these EOM are valid even in the infinite amplitude case of the the expanding cloud of BEC released from its trapping potential in 1D. We also relax the assumption that the components need to be completely overlapping.
	
	Following the method of Ref.~\cite{dalfovo97,kasamatsu04} we present here the EOM in 1D. The 1D hydrodynamical equations [\eq{eq:hydrodynamic1}, \eq{eq:hydrodynamic2}] admit a class of analytical solution with the same form as \eq{eq:rho} but with the ansatz, 
	\begin{gather}
	\rho_l^n (x,t)=\alpha_{l,0}^n (t)-\alpha_l^n (t) x^2~,\label{eq:rhoTF_1D}\\
	v_l^n(x,t) = \beta_l^n (t)x~,\label{eq:vTF_1D}
	\end{gather}
	where $n = o,s$ indicates whether one is in the overlap or singular region. Substitution of \eq{eq:rhoTF_1D} and (\ref{eq:vTF_1D}) into the hydrodynamical equations yield the following constraints:
	\begin{gather}
	-\dot{\alpha}_{l}^n = 3 \alpha_{l}^n \beta_{l}^n~,\label{eq:alphaTF_1D}\\
	-\dot{\beta}_{l}^o = (\beta_{l}^o)^2 + \omega(t)^2 - \frac{2 g_{ll}}{m} \alpha_{l}^o - \frac{2 g_{ll'}}{m} \alpha_{l'}^o~,\label{eq:betaoTF_1D}\\
	-\dot{\beta}_{l}^s = (\beta_{l}^s)^2 + \omega(t)^2 - \frac{2 g_{ll}}{m} \alpha_{l}^s~.\label{eq:betasTF_1D}
	\end{gather}
	These equations are further simplified by introducing adimensional parameter $\lambda_{l}^n$, defined by $\alpha_l^o=\frac{g_{ll}-g_{ll'}}{2|g|(\lambda_l^o)^3}$ and $\alpha_l^s=\frac{1}{2|g|(\lambda_l^s)^3}$. With this substitution, \eq{eq:alphaTF_1D} reduces to $\beta_l^n=\dot{\lambda}_l^n/\lambda_l^n$ and \eq{eq:betaoTF_1D} and \eq{eq:betasTF_1D} become,
	\begin{gather}
	\ddot{\lambda}_{l}^o = \frac{g_{ll}(g_{l'l'} - g_{ll'})}{|g| (\lambda_{l}^o)^2} +  \frac{g_{ll'}(g_{ll} - g_{ll'}) \lambda_{l}^o}{|g| (\lambda_{l}^o)^3} - \omega(t)^2 \lambda_{l}^o~,\label{eq:lambdao1D}\\
	\ddot{\lambda}_{l}^s = (\lambda_l^s)^{-2} - \omega(t)^2 \lambda_{l}^s~.\label{eq:lambdas1D}
	\end{gather}
	\eq{eq:lambdao1D} and (\ref{eq:lambdas1D}) represent the EOM of the two-component BEC in the TF limit in 1D. For completeness a derivation for the general 3D case, in the notation used in this paper, can be found in \ref{app:A}. \eq{eq:lambdao1D} and (\ref{eq:lambdas1D}) form six coupled second-order differential equations for the overlap and singular regions respectively. The last terms of \eq{eq:lambdao1D} and (\ref{eq:lambdas1D}) represent the effects of the confining potential, whereas the other term arises from the particle-particle interactions. In the overlap region the dynamics of the system is dependent on the intra- and inter-component interaction strengths. In the singular region the dynamics of the system are independent of the interaction strengths, as is the case for single component dynamics in the large particle number limit \cite{castin96}. The solutions of \eq{eq:lambdao1D} and (\ref{eq:lambdas1D}) determine $\alpha_{l}^n(t)$. 	
	
	$\alpha_{l,0}^n(t)$ are determined by boundary conditions. We consider the case when $g_{11} < g_{22}$, such that component 1 will never form a singular region; in the TF limit, component 2 will form singular regions surrounding component 1 (note the situation is simply reversed for $g_{11} > g_{22}$).  Application of the conditions of normalization and continuity at the overlap-singular boundary yield (see \ref{app:B}),
	\begin{gather}
	\alpha_{1,0}^o(t) = \bigl[ \frac32 \sqrt{\alpha_1^o(t)}\bigr]^{2/3}~,\label{eq:alpha10o1D}\\
	\alpha_{2,0}^s(t) = \bigl\{\frac{3-12\sqrt{\alpha_1^s(t)} [\alpha_2^o(t) + \alpha_2^s(t)]}{8 \alpha_1^o(t)} \bigl\}^{2/3}~,\label{eq:alpha20s1D}\\
	\alpha_{2,0}^o(t) = \alpha_{2,0}^s(t) + \bigl[ \frac{3}{2 \alpha_1^o(t)} \bigr]^{2/3}[\alpha_2^o(t) + \alpha_2^s(t)]~.\label{eq:alpha20o1D}
	\end{gather}
	The overlap-singular boundary occurs at $R_1 (t)=\sqrt{\alpha_{1,0}^o(t) / \alpha_{1}^o(t)}=\bigl[\frac{3}{2\alpha_1(t)}\bigr]^{1/3}$. The location where the density vanishes (which gives the condensate width) is given by $R_2(t) = \sqrt{\alpha_{2,0}^s(t)/\alpha_{2}^s(t)}$.
	
	The confining potential in the $x$-direction is switched off in our model by setting $\omega(t>0)=0$ in \eq{eq:lambdao1D} and (\ref{eq:lambdas1D}).  We then solve these EOM to predict the dynamics of the released gas.  
	
\begin{figure}[tb]%
		\centering
		\includegraphics[width=\columnwidth]{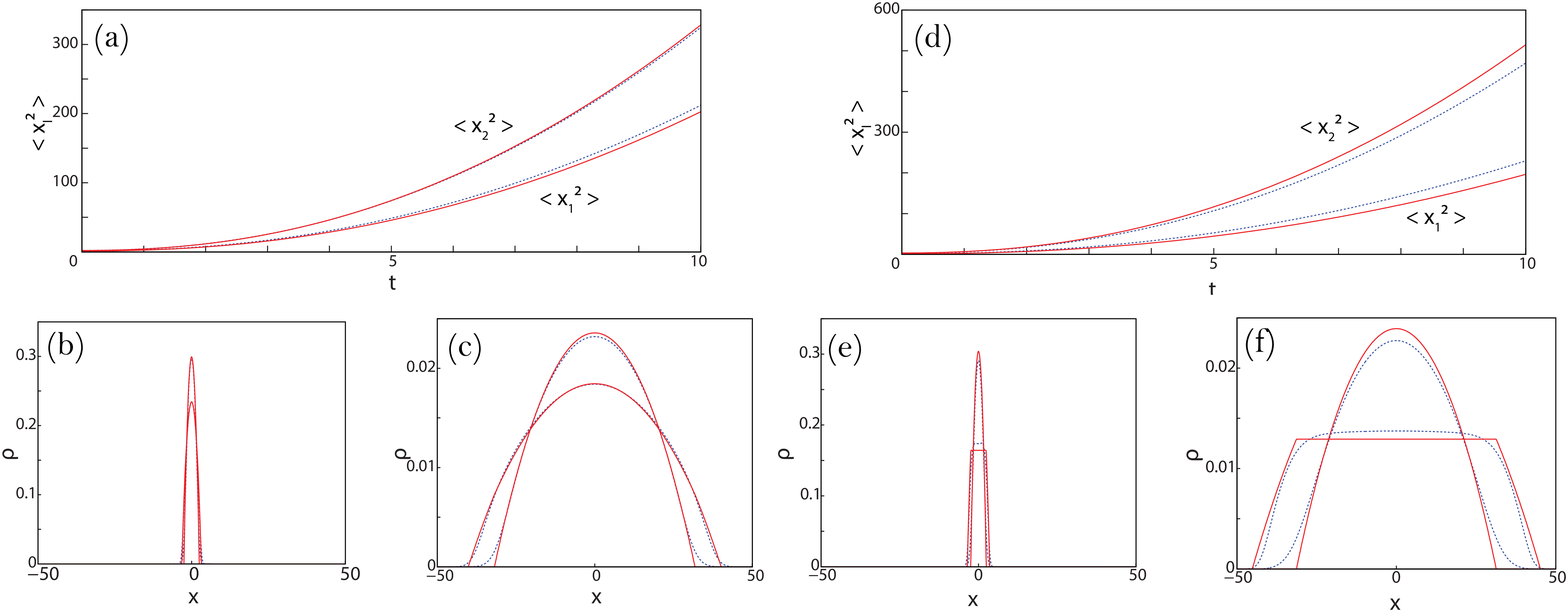}%
		\caption{(Color online) Expansion of a two-component BEC released from a trap for $g_{11}=10$, $g_{22}=20$, (a)-(c)$g_{12}=1$ and (d)-(f)$g_{12}=10$. (a) and (d) show the mean-squared width of the two-components of the BEC over time. (b) and (e) are the initial population distribution of the two-components and (c) and (f) are the population distribution at $t=10$. The solid (red) lines represent the EOM predictions and the dotted (blue) lines the results from computer simulations. The EOM approximate the behaviour of the released BEC well, and becomes more accurate as $g_{12}^2/|g_{11} g_{22}|$ gets smaller.}%
		\label{fig:rel}%
	\end{figure}

	\fig{fig:rel} compares the evolution of the released gas as predicted by the EOM with computer simulations of the full model. The top plots show the mean-square displacement for $g_{12}^2/g_{11} g_{22} = 5\times10^{-3}$ and 0.5. The bottom plots of \fig{fig:rel} show the normalized population distribution at $t=0$ and 10. \fig{fig:rel} shows that the EOM is a good approximation of the expansion of the two-component released gas, being more accurate as one approaches the TF regime, $g_{12} \ll |g_{11}|,|g_{22}|$.
	
\begin{figure}[tb]%
		\centering
		\includegraphics[width=\columnwidth]{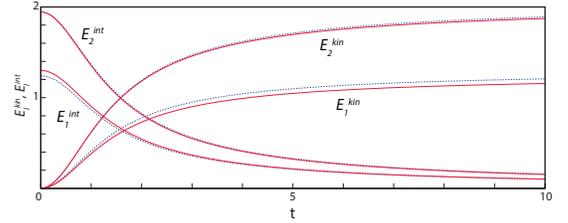}%
		\caption{(Color online) Kinetic $E_l^\text{kin}$ and interaction $E_l^\text{int}$ energy of expanding two-component BEC released from trap for $g_{11}=10, g_{22}=20, g_{12}=1$. The solid (red) lines represent the EOM predictions and the dotted (blue) lines the results from computer simulations. As the cloud expands the interaction energy is converted to kinetic energy. The EOM provides a good approximation of the release energy of the BEC.}%
		\label{fig:ke_int}%
	\end{figure}

	One can also use the EOM to calculate the release energy of the BEC.  Using the solutions of the EOM, \fig{fig:ke_int} plots the kinetic energy and interaction energy of the released BEC in the TF limit, and compares it to the simulated kinetic and interaction energies.
	
\begin{figure}[tb]%
		\centering
		\includegraphics[width=\columnwidth]{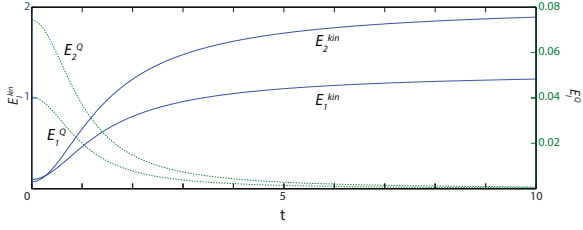}%
		\caption{(Color online) Kinetic ($E_l^\text{kin} = E_l^Q + E_l^\phi$) and quantum pressure energy ($E_l^Q$) of expanding two-component BEC released from a trap. The solid (blue) lines represents $E_l^\text{kin}$ and the dotted (green) lines $E_l^Q$. As $E_l^\text{kin}  \gg E_l^Q$, most of the kinetic energy comes from $E_l^\phi$. Note the different energy scales used for $E_l^\text{kin}$ and $E_l^Q$. Furthermore $E_l^Q$ vanishes as the BEC expands. Scattering matrix $g$ is as in \fig{fig:ke_int}. }%
		\label{fig:ke_qe}%
	\end{figure}

	\fig{fig:ke_int} shows that as the BEC expands the interaction energy is converted to kinetic energy. The calculation of the release energy from the EOM is a good approximation because here $E_l^Q$ is small relative to $E_l^\phi$ and $E_l^\text{int}$, and quickly decreases to $E_l^Q (t \rightarrow \infty) = 0$ as the BEC expands, as shown \fig{fig:ke_qe}. 
	
	\section{Conclusion and outlook}
	
	We have derived an analytical formulation for the beating frequencies seen in two-component BECs. We have also confirmed that the large oscillation amplitude EOM for two-component BECs are valid even in the infinite oscillation amplitude case of the freely expanding condensate.
	The formulation and analysis in this work contributes to a better understandings of the large mulitivariate parameter space that characterises two-component BEC systems. It would be interesting in further work to extend the analysis of the beating frequencies to 3D. This extension however is non-trivial, as the extra dimensions introduce other modes of oscillations, such as the quadrupole and scissor modes. Nevertheless, our work should provide a framework upon which a generalisation can be achieved.   
	
	\section{Acknowledgments}
	 The author would like to thank A . L. C. Hayward for fundamental conceptual discussions, and M. Oshikawa for his support. The author would also like to thank C.-H. Su, R. A. Henry, A. M. Martin, and B. Mulkerin for their feedback on the manuscript. This work was in part supported by the Japan Society for the Promotion of Science.
	
	\appendix
	\section{Generalisation of the equations of motion to 3D}
	\label{app:A}
	In this appendix we generalise the equations of motion for two-component BECs to 3D. Here we provide the general case were the mass and number of atoms of each of the components may differ, and the trapping frequencies in the three spatial directions are free to vary. Note that this derivation is independent of parameter rescaling.
	
	In the TF limit, the 3D hydrodynamical equations for the two-component BEC are
	\begin{gather}
	-\frac{\partial\rho_l}{\partial t} = \nabla \cdot (\mathbf{v}_l \rho_l)~,\label{eq:hydrodynamic1_3D}\\
	-m_l \frac{\partial\mathbf{v}_l}{\partial t} = \nabla(V_l + g_{ll}\rho_l + g_{ll'} \rho_{l'} + \frac12 m_l \mathbf{v}_l^2)~.\label{eq:hydrodynamic_3D}
	\end{gather}
	
	A solution to the 3D hydrodynamical equations has the same form as \eq{eq:rho} but with the ansatz,
	\begin{gather}
	\rho_l^n~(\mathbf{v},t) = \alpha_{l,0}^n(t) - \alpha_{l,x}^n(t) x^2 - \alpha_{l,y}^n(t) y^2 - \alpha_{l,z}^n(t) z^2~,\label{eq:rhoTF}\\
	\mathbf{v}_l^n(\mathbf{r},t) = \frac12 \nabla [\beta_{l,x}^n(t)x^2 + \beta_{l,y}^n(t)y^2 + \beta_{l,z}^n(t)z^2]~.\label{eq:vTF}
	\end{gather}
	where $n = o,s$ indicates whether one is in the overlap or singular region. Substitution of \eq{eq:rhoTF} and (\ref{eq:vTF}) into the hydrodynamical equations yield the following constraints:
	\begin{gather}
	-\dot{\alpha}_{l,i}^n = 2 \alpha_{l,i}^n \beta_{l,i}^n + \alpha_{l,i}^n \sum_j{\beta_{l,j}^n} ~,\label{eq:alphaTF}\\
	-\dot{\beta}_{l,i}^o = (\beta_{l,i}^o)^2 + \omega_{l,i}^2 - \frac{2 g_{ll}}{m_l} \alpha_{l,i}^o - \frac{2 g_{ll'}}{m_l} \alpha_{l',i}^o~,\label{eq:betaoTF}\\
	-\dot{\beta}_{l,i}^s = (\beta_{l,i}^s)^2 + \omega_{l,i}^2 - \frac{2 g_{ll}}{m_l} \alpha_{l,i}^s~.\label{eq:betasTF}
	\end{gather}
	where $i,j=x,y,z$. These equations are further simplified by introducing adimensional parameter $\lambda_{l,i}^n$, defined by $\alpha_{l,i}^o = \frac{m_l \omega_{l,i}^2(0) (g_{ll}- g_{ll'})}{2 |g| \lambda_{l,i}^o \prod_j{\lambda_{l,j}^o} }$ and $\alpha_{l,i}^s = \frac{m_l \omega_{l,i}^2(0)}{2 |g| \lambda_{l,i}^s \prod_j{\lambda_{l,j}^s}}$. With this substitution, \eq{eq:alphaTF} reduces to $\beta_{l,i}^n = \dot{\lambda}_{l,i}^n / \lambda_{l,i}^n$ and \eq{eq:betaoTF} and \eq{eq:betasTF} become,
	\begin{align}
	\ddot{\lambda}_{l,i}^o &= \frac{g_{ll}(g_{l'l'} - g_{ll'}) \omega_l(0)^2}{|g| \lambda_{l,i}^o \prod_j{\lambda_{l,j}^o}}\\
	&+  \frac{g_{ll'}(g_{ll} - g_{ll'}) m_{l'} \omega_{l'}(0)^2 \lambda_{l,i}^o}{|g| m_{l'} (\lambda_{l,i}^o)^2 \prod_j{\lambda_{l',j}^o} } - \omega_l(t)^2 \lambda_{l,i}^o~,\label{eq:lambdao}\\
	\ddot{\lambda}_{l,i}^s &= \frac{\omega_l(0)^2}{|g| \lambda_{l,i}^s \prod_j{\lambda_{l,j}^s}} - \omega_l(t)^2 \lambda_{l,i}^s~.\label{eq:lambdas}
	\end{align}
	\eq{eq:lambdao} and (\ref{eq:lambdas}) represent the EOM of the two-component BEC in the TF limit. Their solutions determine $\alpha_{l,i}^n(t)$. $\alpha_{l,0}^n(t)$ is determined by conditions of continuity at the overlap-singular boundary  $\rho_l^o (\mathbf{R})=\rho_{l'}^s(\mathbf{R})$ (where $\mathbf{R}$ is locates the boundary), and normalization, $N_l=\int\rho_l(\mathbf{r}) d\mathbf{r}$.
	
	\section{Overlap-singular boundary conditions for 1D}
	\label{app:B}
	Applying the normalisation condition ($\int|\psi_l|^2 dx = 1$ ) to component 1,
	\begin{equation}
	\int_{-R_1(t)}^{R_1(t)} \alpha_{1,0}^o (t)-\alpha_1^o (t) x^2 dx = 1~,
	\end{equation}
	we solve for $\alpha_{1,0}^o(t)$,
	\begin{equation}
	\alpha_{1,0}^o(t) = \bigl[ \frac32 \sqrt{\alpha_1^o(t)}\bigr]^{2/3}~.
	\end{equation}
	
	Similarly for component 2,
	\begin{equation}
	\int_{0}^{R_1(t)} \alpha_{2,0}^o (t)-\alpha_2^o (t) x^2 dx + \int_{R_1(t)}^{R_2(t)} \alpha_{2,0}^s (t)-\alpha_2^s (t) x^2 dx= 1/2~,
	\end{equation}
	we get
	\begin{equation}
	\alpha_{2,0}^s(t) = \bigl\{\frac{3-12\sqrt{\alpha_1^s(t)} [\alpha_2^o(t) + \alpha_2^s(t)]}{8 \alpha_1^o(t)} \bigl\}^{2/3}~.
	\end{equation}
	
	Finally we apply the continuity condition, $\rho_1^o(R_1,t)=\rho_2^o(R_1,t)$, to get,
	\begin{equation}
	\alpha_{2,0}^o(t) = \alpha_{2,0}^s(t) + \bigl[ \frac{3}{2 \alpha_1^o(t)} \bigr]^{2/3}[\alpha_2^o(t) + \alpha_2^s(t)]~.
	\end{equation}
	
	Note that the overlap-singular boundary (where $\rho_1^0=0$) occurs at $R_1 (t)=\sqrt{\alpha_{1,0}^o(t) / \alpha_{1}^o(t)}=\bigl[\frac{3}{2\alpha_1(t)}\bigr]^{1/3}$. Similarly, $R_2(t) = \sqrt{\alpha_{2,0}^s(t)/\alpha_{2}^s(t)}$.
	
	\section*{References}
	\bibliographystyle{unsrt}
	\bibliography{bubble_jpb}

\begin{thebibliography}{10}

\bibitem{myatt97}
C.~J. Myatt, E.~A. Burt, R.~W. Ghrist, E.~A. Cornell, and C.~E. Wieman.
\newblock Production of two overlapping bose-einstein condensates by
  sympathetic cooling.
\newblock {\em Phys. Rev. Lett.}, 78:586--589, Jan 1997.

\bibitem{hall98}
D.~S. Hall, M.~R. Matthews, J.~R. Ensher, C.~E. Wieman, and E.~A. Cornell.
\newblock Dynamics of component separation in a binary mixture of bose-einstein
  condensates.
\newblock {\em Phys. Rev. Lett.}, 81:1539--1542, Aug 1998.

\bibitem{maddaloni00}
P.~Maddaloni, M.~Modugno, C.~Fort, F.~Minardi, and M.~Inguscio.
\newblock Collective oscillations of two colliding bose-einstein condensates.
\newblock {\em Phys. Rev. Lett.}, 85:2413--2417, Sep 2000.

\bibitem{delannoy01}
G.~Delannoy, S.~G. Murdoch, V.~Boyer, V.~Josse, P.~Bouyer, and A.~Aspect.
\newblock Understanding the production of dual bose-einstein condensation with
  sympathetic cooling.
\newblock {\em Phys. Rev. A}, 63:051602, Apr 2001.

\bibitem{mertes07}
K.~M. Mertes, J.~W. Merrill, R.~Carretero-Gonz\'alez, D.~J. Frantzeskakis,
  P.~G. Kevrekidis, and D.~S. Hall.
\newblock Nonequilibrium dynamics and superfluid ring excitations in binary
  bose-einstein condensates.
\newblock {\em Phys. Rev. Lett.}, 99:190402, Nov 2007.

\bibitem{anderson09}
R.~P. Anderson, C.~Ticknor, A.~I. Sidorov, and B.~V. Hall.
\newblock Spatially inhomogeneous phase evolution of a two-component
  bose-einstein condensate.
\newblock {\em Phys. Rev. A}, 80:023603, Aug 2009.

\bibitem{tojo10}
Satoshi Tojo, Yoshihisa Taguchi, Yuta Masuyama, Taro Hayashi, Hiroki Saito, and
  Takuya Hirano.
\newblock Controlling phase separation of binary bose-einstein condensates via
  mixed-spin-channel feshbach resonance.
\newblock {\em Phys. Rev. A}, 82:033609, Sep 2010.

\bibitem{ferrari02}
G.~Ferrari, M.~Inguscio, W.~Jastrzebski, G.~Modugno, G.~Roati, and A.~Simoni.
\newblock Collisional properties of ultracold k-rb mixtures.
\newblock {\em Phys. Rev. Lett.}, 89:053202, Jul 2002.

\bibitem{modugno02}
G.~Modugno, M.~Modugno, F.~Riboli, G.~Roati, and M.~Inguscio.
\newblock Two atomic species superfluid.
\newblock {\em Phys. Rev. Lett.}, 89:190404, Oct 2002.

\bibitem{thalhammer08}
G.~Thalhammer, G.~Barontini, L.~De~Sarlo, J.~Catani, F.~Minardi, and
  M.~Inguscio.
\newblock Double species bose-einstein condensate with tunable interspecies
  interactions.
\newblock {\em Phys. Rev. Lett.}, 100:210402, May 2008.

\bibitem{mccarron11}
D.~J. McCarron, H.~W. Cho, D.~L. Jenkin, M.~P. K\"oppinger, and S.~L. Cornish.
\newblock Dual-species bose-einstein condensate of $^{87}\mathrm{Rb}$ and
  $^{133}\mathrm{Cs}$.
\newblock {\em Phys. Rev. A}, 84:011603, Jul 2011.

\bibitem{papp08}
S.~B. Papp, J.~M. Pino, and C.~E. Wieman.
\newblock Tunable miscibility in a dual-species bose-einstein condensate.
\newblock {\em Phys. Rev. Lett.}, 101:040402, Jul 2008.

\bibitem{robins01}
Nicholas~P. Robins, Weiping Zhang, Elena~A. Ostrovskaya, and Yuri~S. Kivshar.
\newblock Modulational instability of spinor condensates.
\newblock {\em Phys. Rev. A}, 64:021601, Jul 2001.

\bibitem{ronen08}
Shai Ronen, John~L. Bohn, Laura~Elisa Halmo, and Mark Edwards.
\newblock Dynamical pattern formation during growth of a dual-species
  bose-einstein condensate.
\newblock {\em Phys. Rev. A}, 78:053613, Nov 2008.

\bibitem{busch01}
Th. Busch and J.~R. Anglin.
\newblock Dark-bright solitons in inhomogeneous bose-einstein condensates.
\newblock {\em Phys. Rev. Lett.}, 87:010401, Jun 2001.

\bibitem{hamner11}
C.~Hamner, J.~J. Chang, P.~Engels, and M.~A. Hoefer.
\newblock Generation of dark-bright soliton trains in superfluid-superfluid
  counterflow.
\newblock {\em Phys. Rev. Lett.}, 106:065302, Feb 2011.

\bibitem{matthews99}
M.~R. Matthews, B.~P. Anderson, P.~C. Haljan, D.~S. Hall, C.~E. Wieman, and
  E.~A. Cornell.
\newblock Vortices in a bose-einstein condensate.
\newblock {\em Phys. Rev. Lett.}, 83:2498--2501, Sep 1999.

\bibitem{feder99}
David~L. Feder, Charles~W. Clark, and Barry~I. Schneider.
\newblock Vortex stability of interacting bose-einstein condensates confined in
  anisotropic harmonic traps.
\newblock {\em Phys. Rev. Lett.}, 82:4956--4959, Jun 1999.

\bibitem{kasamatsu04}
Kenichi Kasamatsu, Makoto Tsubota, and Masahito Ueda.
\newblock Quadrupole and scissors modes and nonlinear mode coupling in trapped
  two-component bose-einstein condensates.
\newblock {\em Phys. Rev. A}, 69:043621, Apr 2004.

\bibitem{hoefer11}
M.~A. Hoefer, J.~J. Chang, C.~Hamner, and P.~Engels.
\newblock Dark-dark solitons and modulational instability in miscible
  two-component bose-einstein condensates.
\newblock {\em Phys. Rev. A}, 84:041605, Oct 2011.

\bibitem{graham98}
Robert Graham and Dan Walls.
\newblock Collective excitations of trapped binary mixtures of bose-einstein
  condensed gases.
\newblock {\em Phys. Rev. A}, 57:484--487, Jan 1998.

\bibitem{ho96}
Tin-Lun Ho and V.~B. Shenoy.
\newblock Binary mixtures of bose condensates of alkali atoms.
\newblock {\em Phys. Rev. Lett.}, 77:3276--3279, Oct 1996.

\bibitem{gorlitz01}
A.~G\"orlitz, J.~M. Vogels, A.~E. Leanhardt, C.~Raman, T.~L. Gustavson, J.~R.
  Abo-Shaeer, A.~P. Chikkatur, S.~Gupta, S.~Inouye, T.~Rosenband, and
  W.~Ketterle.
\newblock Realization of bose-einstein condensates in lower dimensions.
\newblock {\em Phys. Rev. Lett.}, 87:130402, Sep 2001.

\bibitem{moritz03}
Henning Moritz, Thilo St\"oferle, Michael K\"ohl, and Tilman Esslinger.
\newblock Exciting collective oscillations in a trapped 1d gas.
\newblock {\em Phys. Rev. Lett.}, 91:250402, Dec 2003.

\bibitem{haller09}
Elmar Haller, Mattias Gustavsson, Manfred~J. Mark, Johann~G. Danzl, Russell
  Hart, Guido Pupillo, and Hanns-Christoph Nägerl.
\newblock Realization of an excited, strongly correlated quantum gas phase.
\newblock {\em Science}, 325(5945):1224--1227, 2009.

\bibitem{cerimele00}
M.~M. Cerimele, M.~L. Chiofalo, F.~Pistella, S.~Succi, and M.~P. Tosi.
\newblock Numerical solution of the gross-pitaevskii equation using an explicit
  finite-difference scheme:\quad{}an application to trapped bose-einstein
  condensates.
\newblock {\em Phys. Rev. E}, 62:1382--1389, Jul 2000.

\bibitem{bao03}
Weizhu Bao and Weijun Tang.
\newblock Ground-state solution of bose-einstein condensate by directly
  minimizing the energy functional.
\newblock {\em Journal of Computational Physics}, 187(1):230 -- 254, 2003.

\bibitem{jackson98}
A.~D. Jackson, G.~M. Kavoulakis, and C.~J. Pethick.
\newblock Solitary waves in clouds of bose-einstein condensed atoms.
\newblock {\em Phys. Rev. A}, 58:2417--2422, Sep 1998.

\bibitem{dunjko01}
V.~Dunjko, V.~Lorent, and M.~Olshanii.
\newblock Bosons in cigar-shaped traps: Thomas-fermi regime, tonks-girardeau
  regime, and in between.
\newblock {\em Phys. Rev. Lett.}, 86:5413--5416, Jun 2001.

\bibitem{ge05}
Ge~Yunyi.
\newblock {\em Dimension reduction of the Gross-Pitaevskii equation for
  Bose-Einstein condensates}.
\newblock PhD thesis, 2005.

\bibitem{baym96}
Gordon Baym and C.~J. Pethick.
\newblock Ground-state properties of magnetically trapped bose-condensed
  rubidium gas.
\newblock {\em Phys. Rev. Lett.}, 76:6--9, Jan 1996.

\bibitem{corro09}
I.~Corro, R.~G. Scott, and A.~M. Martin.
\newblock Dynamics of two-component bose-einstein condensates in rotating
  traps.
\newblock {\em Phys. Rev. A}, 80:033609, Sep 2009.

\bibitem{trippenbach00}
Marek Trippenbach, Krzysztof Góral, Kazimierz Rzazewski, Boris Malomed, and
  Y~B Band.
\newblock Structure of binary bose-einstein condensates.
\newblock {\em Journal of Physics B: Atomic, Molecular and Optical Physics},
  33(19):4017, 2000.

\bibitem{stringari96}
S.~Stringari.
\newblock Collective excitations of a trapped bose-condensed gas.
\newblock {\em Phys. Rev. Lett.}, 77:2360--2363, Sep 1996.

\bibitem{jin96}
D.~S. Jin, J.~R. Ensher, M.~R. Matthews, C.~E. Wieman, and E.~A. Cornell.
\newblock Collective excitations of a bose-einstein condensate in a dilute gas.
\newblock {\em Phys. Rev. Lett.}, 77:420--423, Jul 1996.

\bibitem{mewes96}
M.-O. Mewes, M.~R. Andrews, N.~J. van Druten, D.~M. Kurn, D.~S. Durfee, C.~G.
  Townsend, and W.~Ketterle.
\newblock Collective excitations of a bose-einstein condensate in a magnetic
  trap.
\newblock {\em Phys. Rev. Lett.}, 77:988--991, Aug 1996.

\bibitem{castin96}
Y.~Castin and R.~Dum.
\newblock Bose-einstein condensates in time dependent traps.
\newblock {\em Phys. Rev. Lett.}, 77:5315--5319, Dec 1996.

\bibitem{dalfovo97}
F.~Dalfovo, C.~Minniti, and L.~P. Pitaevskii.
\newblock Frequency shift and mode coupling in the nonlinear dynamics of a
  bose-condensed gas.
\newblock {\em Phys. Rev. A}, 56:4855--4863, Dec 1997.

\bibitem{garcia99a}
Juan~J. Garc\'{i}a-Ripoll, V\'{i}ctor~M. P\'{e}rez-Garc\'{i}a, and Pedro
  Torres.
\newblock Extended parametric resonances in nonlinear schrödinger systems.
\newblock {\em Phys. Rev. Lett.}, 83:1715--1718, Aug 1999.

\bibitem{garcia99b}
Juan J.~G. Ripoll and V\'{i}ctor~M. P\'erez-Garc\'{i}a.
\newblock Barrier resonances in bose-einstein condensation.
\newblock {\em Phys. Rev. A}, 59:2220--2231, Mar 1999.

\bibitem{pollack09}
S.~E. Pollack, D.~Dries, M.~Junker, Y.~P. Chen, T.~A. Corcovilos, and R.~G.
  Hulet.
\newblock Extreme tunability of interactions in a $^{7}\mathrm{Li}$
  bose-einstein condensate.
\newblock {\em Phys. Rev. Lett.}, 102:090402, Mar 2009.

\bibitem{vidanovic11}
Ivana Vidanovi\ifmmode~\acute{c}\else \'{c}\fi{}, Antun
  Bala\ifmmode~\check{z}\else \v{z}\fi{}, Hamid Al-Jibbouri, and Axel Pelster.
\newblock Nonlinear bose-einstein-condensate dynamics induced by a harmonic
  modulation of the $s$-wave scattering length.
\newblock {\em Phys. Rev. A}, 84:013618, Jul 2011.

\bibitem{busch97}
Th. Busch, J.~I. Cirac, V.~M. Perez-Garcia, and P.~Zoller.
\newblock Stability and collective excitations of a two-component bose-einstein
  condensed gas: A moment approach.
\newblock {\em Phys. Rev. A}, 56:2978--2983, Oct 1997.

\bibitem{esry98}
B.~D. Esry and Chris~H. Greene.
\newblock Low-lying excitations of double bose-einstein condensates.
\newblock {\em Phys. Rev. A}, 57:1265--1271, Feb 1998.

\bibitem{pu98}
H.~Pu and N.~P. Bigelow.
\newblock Collective excitations, metastability, and nonlinear response of a
  trapped two-species bose-einstein condensate.
\newblock {\em Phys. Rev. Lett.}, 80:1134--1137, Feb 1998.

\bibitem{gordon98}
D.~Gordon and C.~M. Savage.
\newblock Excitation spectrum and instability of a two-species bose-einstein
  condensate.
\newblock {\em Phys. Rev. A}, 58:1440--1444, Aug 1998.

\end{thebibliography}
	
\end{document}